\documentclass[%
 reprint,superscriptaddress,
 amsmath,amssymb,prx,longbibliography]{revtex4-1}

\usepackage{physics}
\usepackage{xcolor}
\usepackage{bm}
\usepackage{xtab,afterpage,longtable}
\usepackage{lipsum}
\usepackage{graphicx}
\usepackage{dcolumn}
\usepackage{booktabs} 
\usepackage{multirow}
\usepackage{color}
\usepackage{eucal}
\usepackage{mathrsfs}  
\usepackage{siunitx}
\usepackage{makecell}
\usepackage{hyperref}

\begin{document}
\title{Comment on ``High-Power Collective Charging of a Solid-State Quantum Battery''}

\author{Haowei Xu}%
\email{haoweixu@mit.edu}
\affiliation{Department of Nuclear Science and Engineering, Massachusetts Institute of Technology, Cambridge, MA 02139, USA}

\author{Ju Li}%
 \email{liju@mit.edu}
 \affiliation{Department of Nuclear Science and Engineering, Massachusetts Institute of Technology, Cambridge, MA 02139, USA}
 \affiliation{Department of Materials Science and Engineering, Massachusetts Institute of Technology, MA 02139, USA}

\maketitle

In the Letter~\cite{ferraro2018high},  Ferraro \emph{et al.} claimed a quantum advantage in the Dicke quantum battery (QB), whereby $N$ two-level systems (TLS) are coupled to a common photonic mode of a cavity. They argued that compared with the so-called Rabi QB, the Dicke QB exhibits a $\sqrt{N}$ quantum enhancement in the charging power because of the entanglement created by the common photonic mode. In this Comment, however, we demonstrate that the apparent $\sqrt{N}$ enhancement actually comes from the stronger cavity electric (or magnetic) field under the setup discussed in Ref.~\cite{ferraro2018high}. This is a trivial classical effect, and there is no true ``quantum advantage" in the charging power of the Dicke QB. While somewhat similar questions regarding the origin of the claimed ``quantum advantage" have been raised in Ref.~\cite{andolina2019quantum,julia2020bounds,quach2022superabsorption}, here we would like to make it clear that the $\sqrt{N}$ enhancement in Ref.~\cite{ferraro2018high} is purely a classical effect,  not attributable to quantum entanglement or collective phenomena. Therefore, we believe that using the term  ``quantum battery" in this context may be inappropriate and misleading. 

\begin{figure}
    \centering
    \includegraphics[width=1\linewidth]{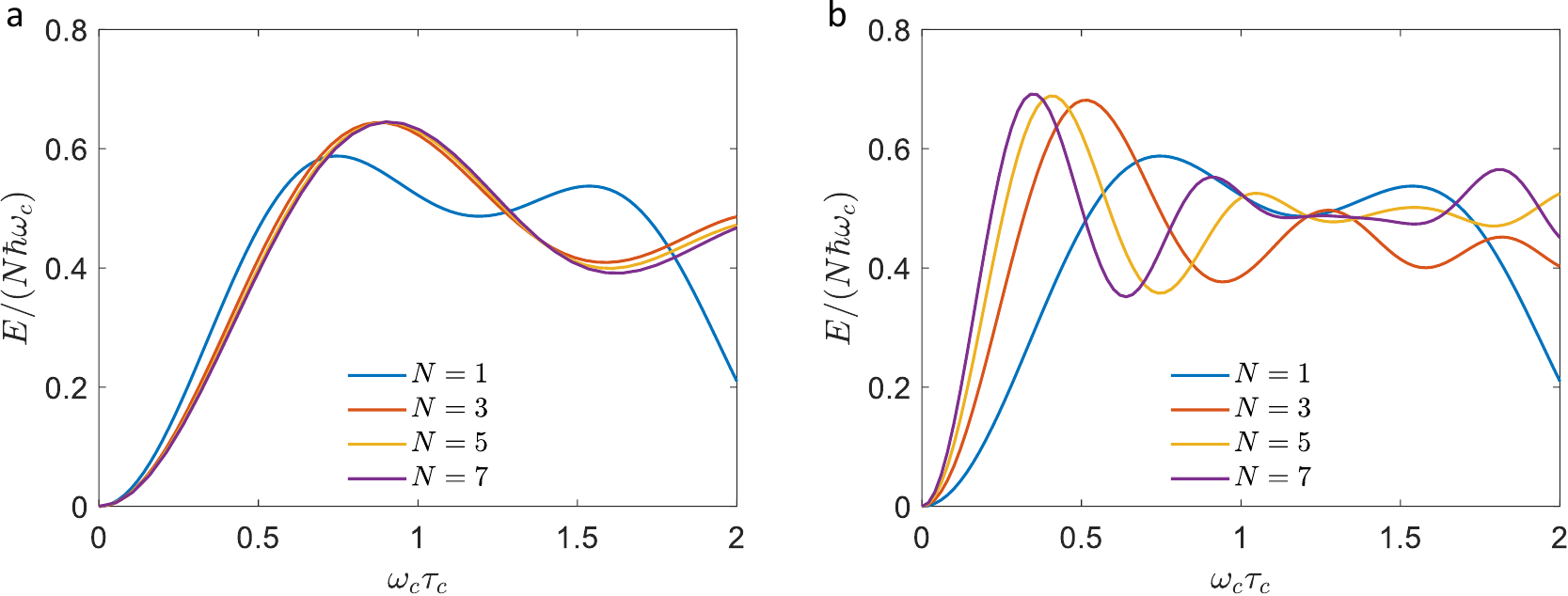}
    \caption{The dependence of stored energy $E$ on charging time $\tau_c$. (a) The TLS-photon coupling strength is $\bar{\lambda}_{\rm Dicke} = \bar{\lambda}/\sqrt{N}$, so that the initial cavity field strength $\mathcal{F}$ is independent of $N$. The maximum charging power is almost the same for different $N$. (b) The TLS-photon coupling strength is $\bar{\lambda}_{\rm Dicke} = \bar{\lambda}$, which is considered in Ref.~\cite{ferraro2018high}. In this case, one has $\mathcal{F}\propto \sqrt{N}$, and the charging power is higher for larger $N$. This, however, is not a quantum advantage.}
    \label{fig:quantum_battery}
\end{figure}

We start with Eqs.~(1, 2) and Fig.~(1) in Ref.~\cite{ferraro2018high}, which describes the basis setup of the Dicke QB. Notably,  the authors assumed that the photon-TLS coupling strength $\bar{\lambda}$  is the same for both Rabi and Dicke QB. However, $\bar{\lambda}$ should not be considered as a constant. It actually depends on the mode volume $V$ of the cavity, that is, one has $\bar{\lambda}\propto \mathcal{F}_{\rm zpf} \propto \sqrt{1/V}$, where $\mathcal{F}_{\rm zpf}$ is the zero-point field of the cavity.  In Fig.~(1) of Ref.~\cite{ferraro2018high}, the illustration seems to suggest that the Dicke QB considered by the authors can be realized by simply removing the mirrors in between single Rabi QBs. In this case, one would have $V_{\rm Dicke} = NV_{\rm Rabi}$ and $\bar{\lambda}_{\rm Dicke} =  \bar{\lambda}_{\rm Rabi}/\sqrt{N}$, where the subscripts denote the type of QB. This would eliminate the $\sqrt{N}$ quantum advantage claimed by the authors (Figs.~2 and~3 in Ref.~\cite{ferraro2018high}) . To verify this point, we performed the same simulations as in Ref.~\cite{ferraro2018high}, but with $\bar{\lambda}_{\rm Dicke} =  \bar{\lambda}/\sqrt{N}$. The results are shown in Fig.~\ref{fig:quantum_battery}(a). One can see that the charging power is not improved by large $N$.

The $\sqrt{N}$  advantage claimed by the authors only appears when we use $\bar{\lambda}_{\rm Dicke} = \bar{\lambda}_{\rm Rabi} = \bar{\lambda}$, which requires $V_{\rm Dicke} = V_{\rm Rabi}$ and thus higher TLS density in the Dicke QB. In this case, since initially there are $N$ photons in the Dicke QB [see discussions around Eq.~(1) in Ref.~\cite{ferraro2018high}] , the cavity field strength would be $\mathcal{F} = \sqrt{N} \mathcal{F}_{\rm zpf}$, which is $\sqrt{N}$ times stronger than that of the Rabi QB. We reproduced the charging dynamics under this setup in Fig.~\ref{fig:quantum_battery}(b), and it indeed shows the $\sqrt{N}$ enhancement. However, this enhancement just comes from the larger $\mathcal{F}$, instead of any entanglement among the TLS. If we shrink $V_{\rm Rabi}$ by a factor of $N$, then the cavity field strength and thus the charging power would also increase by a factor of $\sqrt{N}$ in a Rabi QB. Clearly, this is a trivial classical effect, instead of a true quantum advantage.

If we fix the mode volume and number of photons in the cavity, then the total charging power only scales linearly with the number of TLS. This can be better observed by considering a Dicke QB charged by a classical field with strength $\mathcal{F}$. In this case, the total Hamiltonian is [cf. Eq.~(2) in Ref.~\cite{ferraro2018high}] $\mathcal{H} = \sum_{i=1}^{N}\mathcal{H}_i  = \sum_{i=1}^N  \left( \frac{1}{2}\hbar \omega_a \hat{\sigma}^{z}_{i} +  \mathcal{F} d \hat{\sigma}^x_{i}\right)$, where $d$ is the transition dipole, and $\mathcal{H}_i$ is the Hamiltonian for the $i$-th TLS.  Here we assume $\mathcal{F}$ is a strong classical field, so that it is not influenced by the charging process, and its dynamics can thus be ignored. This can be realized by putting a large number of cavity photons in the Dicke QB.  Starting from the ground state $\ket{\Psi(0)} = \ket{g_1,g_2, \cdots g_N }$, the wavefunction of the TLS is $\ket{\Psi(t)} = e^{-i \mathcal{H}t/\hbar} \ket{\Psi(0)} = \otimes_{i=1}^N e^{-i \mathcal{H}_i t/\hbar} \ket{g_i}$, which remains separable for all TLS at all time $t$. Hence, the TLS are independent of one another during the charging process, and the total charging power is simply $N$ times of that in the case of  $N=1$. There is no additional $\sqrt{N}$ enhancement.

It is also fruitful to compare the Dicke QB with Dicke super-radiance (SR)~\cite{dicke1954coherence}. In Dicke QB, the TLS undergo coherent Rabi-like oscillations (see Figure~\ref{fig:quantum_battery}). In contrast, in Dicke SR, the TLS collectively emit or absorb photons through a measurement-like incoherent process, which directly generates entanglement among TLS. This is similar to many quantum algorithms, where measurements are essential~\cite{nielsen2010quantum}. The quantum advantage may only become apparent in the presence of some measurement-like incoherent processes. Drawing analogy to batteries, the quantum advantage may appear in incoherent emission/absorption rates, i.e. the ``interfacial impedance" of charging/discharging the QB.


\bibliography{bibliography}

\begin{thebibliography}{6}%
\makeatletter
\providecommand \@ifxundefined [1]{%
 \@ifx{#1\undefined}
}%
\providecommand \@ifnum [1]{%
 \ifnum #1\expandafter \@firstoftwo
 \else \expandafter \@secondoftwo
 \fi
}%
\providecommand \@ifx [1]{%
 \ifx #1\expandafter \@firstoftwo
 \else \expandafter \@secondoftwo
 \fi
}%
\providecommand \natexlab [1]{#1}%
\providecommand \enquote  [1]{``#1''}%
\providecommand \bibnamefont  [1]{#1}%
\providecommand \bibfnamefont [1]{#1}%
\providecommand \citenamefont [1]{#1}%
\providecommand \href@noop [0]{\@secondoftwo}%
\providecommand \href [0]{\begingroup \@sanitize@url \@href}%
\providecommand \@href[1]{\@@startlink{#1}\@@href}%
\providecommand \@@href[1]{\endgroup#1\@@endlink}%
\providecommand \@sanitize@url [0]{\catcode `\\12\catcode `\$12\catcode `\&12\catcode `\#12\catcode `\^12\catcode `\_12\catcode `\%12\relax}%
\providecommand \@@startlink[1]{}%
\providecommand \@@endlink[0]{}%
\providecommand \url  [0]{\begingroup\@sanitize@url \@url }%
\providecommand \@url [1]{\endgroup\@href {#1}{\urlprefix }}%
\providecommand \urlprefix  [0]{URL }%
\providecommand \Eprint [0]{\href }%
\providecommand \doibase [0]{http://dx.doi.org/}%
\providecommand \selectlanguage [0]{\@gobble}%
\providecommand \bibinfo  [0]{\@secondoftwo}%
\providecommand \bibfield  [0]{\@secondoftwo}%
\providecommand \translation [1]{[#1]}%
\providecommand \BibitemOpen [0]{}%
\providecommand \bibitemStop [0]{}%
\providecommand \bibitemNoStop [0]{.\EOS\space}%
\providecommand \EOS [0]{\spacefactor3000\relax}%
\providecommand \BibitemShut  [1]{\csname bibitem#1\endcsname}%
\let\auto@bib@innerbib\@empty
\bibitem [{\citenamefont {Ferraro}\ \emph {et~al.}(2018)\citenamefont {Ferraro}, \citenamefont {Campisi}, \citenamefont {Andolina}, \citenamefont {Pellegrini},\ and\ \citenamefont {Polini}}]{ferraro2018high}%
  \BibitemOpen
  \bibfield  {author} {\bibinfo {author} {\bibfnamefont {Dario}\ \bibnamefont {Ferraro}}, \bibinfo {author} {\bibfnamefont {Michele}\ \bibnamefont {Campisi}}, \bibinfo {author} {\bibfnamefont {Gian~Marcello}\ \bibnamefont {Andolina}}, \bibinfo {author} {\bibfnamefont {Vittorio}\ \bibnamefont {Pellegrini}}, \ and\ \bibinfo {author} {\bibfnamefont {Marco}\ \bibnamefont {Polini}},\ }\bibfield  {title} {\enquote {\bibinfo {title} {High-power collective charging of a solid-state quantum battery},}\ }\href@noop {} {\bibfield  {journal} {\bibinfo  {journal} {Physical Review Letters}\ }\textbf {\bibinfo {volume} {120}},\ \bibinfo {pages} {117702} (\bibinfo {year} {2018})}\BibitemShut {NoStop}%
\bibitem [{\citenamefont {Andolina}\ \emph {et~al.}(2019)\citenamefont {Andolina}, \citenamefont {Keck}, \citenamefont {Mari}, \citenamefont {Giovannetti},\ and\ \citenamefont {Polini}}]{andolina2019quantum}%
  \BibitemOpen
  \bibfield  {author} {\bibinfo {author} {\bibfnamefont {Gian~Marcello}\ \bibnamefont {Andolina}}, \bibinfo {author} {\bibfnamefont {Maximilian}\ \bibnamefont {Keck}}, \bibinfo {author} {\bibfnamefont {Andrea}\ \bibnamefont {Mari}}, \bibinfo {author} {\bibfnamefont {Vittorio}\ \bibnamefont {Giovannetti}}, \ and\ \bibinfo {author} {\bibfnamefont {Marco}\ \bibnamefont {Polini}},\ }\bibfield  {title} {\enquote {\bibinfo {title} {Quantum versus classical many-body batteries},}\ }\href@noop {} {\bibfield  {journal} {\bibinfo  {journal} {Physical Review B}\ }\textbf {\bibinfo {volume} {99}},\ \bibinfo {pages} {205437} (\bibinfo {year} {2019})}\BibitemShut {NoStop}%
\bibitem [{\citenamefont {Juli{\`a}-Farr{\'e}}\ \emph {et~al.}(2020)\citenamefont {Juli{\`a}-Farr{\'e}}, \citenamefont {Salamon}, \citenamefont {Riera}, \citenamefont {Bera},\ and\ \citenamefont {Lewenstein}}]{julia2020bounds}%
  \BibitemOpen
  \bibfield  {author} {\bibinfo {author} {\bibfnamefont {Sergi}\ \bibnamefont {Juli{\`a}-Farr{\'e}}}, \bibinfo {author} {\bibfnamefont {Tymoteusz}\ \bibnamefont {Salamon}}, \bibinfo {author} {\bibfnamefont {Arnau}\ \bibnamefont {Riera}}, \bibinfo {author} {\bibfnamefont {Manabendra~N}\ \bibnamefont {Bera}}, \ and\ \bibinfo {author} {\bibfnamefont {Maciej}\ \bibnamefont {Lewenstein}},\ }\bibfield  {title} {\enquote {\bibinfo {title} {Bounds on the capacity and power of quantum batteries},}\ }\href@noop {} {\bibfield  {journal} {\bibinfo  {journal} {Physical Review Research}\ }\textbf {\bibinfo {volume} {2}},\ \bibinfo {pages} {023113} (\bibinfo {year} {2020})}\BibitemShut {NoStop}%
\bibitem [{\citenamefont {Quach}\ \emph {et~al.}(2022)\citenamefont {Quach}, \citenamefont {McGhee}, \citenamefont {Ganzer}, \citenamefont {Rouse}, \citenamefont {Lovett}, \citenamefont {Gauger}, \citenamefont {Keeling}, \citenamefont {Cerullo}, \citenamefont {Lidzey},\ and\ \citenamefont {Virgili}}]{quach2022superabsorption}%
  \BibitemOpen
  \bibfield  {author} {\bibinfo {author} {\bibfnamefont {James~Q}\ \bibnamefont {Quach}}, \bibinfo {author} {\bibfnamefont {Kirsty~E}\ \bibnamefont {McGhee}}, \bibinfo {author} {\bibfnamefont {Lucia}\ \bibnamefont {Ganzer}}, \bibinfo {author} {\bibfnamefont {Dominic~M}\ \bibnamefont {Rouse}}, \bibinfo {author} {\bibfnamefont {Brendon~W}\ \bibnamefont {Lovett}}, \bibinfo {author} {\bibfnamefont {Erik~M}\ \bibnamefont {Gauger}}, \bibinfo {author} {\bibfnamefont {Jonathan}\ \bibnamefont {Keeling}}, \bibinfo {author} {\bibfnamefont {Giulio}\ \bibnamefont {Cerullo}}, \bibinfo {author} {\bibfnamefont {David~G}\ \bibnamefont {Lidzey}}, \ and\ \bibinfo {author} {\bibfnamefont {Tersilla}\ \bibnamefont {Virgili}},\ }\bibfield  {title} {\enquote {\bibinfo {title} {Superabsorption in an organic microcavity: Toward a quantum battery},}\ }\href@noop {} {\bibfield  {journal} {\bibinfo  {journal} {Science advances}\ }\textbf {\bibinfo {volume} {8}},\ \bibinfo {pages} {eabk3160} (\bibinfo {year} {2022})}\BibitemShut {NoStop}%
\bibitem [{\citenamefont {Dicke}(1954)}]{dicke1954coherence}%
  \BibitemOpen
  \bibfield  {author} {\bibinfo {author} {\bibfnamefont {Robert~H}\ \bibnamefont {Dicke}},\ }\bibfield  {title} {\enquote {\bibinfo {title} {Coherence in spontaneous radiation processes},}\ }\href@noop {} {\bibfield  {journal} {\bibinfo  {journal} {Physical Review}\ }\textbf {\bibinfo {volume} {93}},\ \bibinfo {pages} {99} (\bibinfo {year} {1954})}\BibitemShut {NoStop}%
\bibitem [{\citenamefont {Nielsen}\ and\ \citenamefont {Chuang}(2010)}]{nielsen2010quantum}%
  \BibitemOpen
  \bibfield  {author} {\bibinfo {author} {\bibfnamefont {Michael~A}\ \bibnamefont {Nielsen}}\ and\ \bibinfo {author} {\bibfnamefont {Isaac~L}\ \bibnamefont {Chuang}},\ }\href@noop {} {\emph {\bibinfo {title} {Quantum computation and quantum information}}}\ (\bibinfo  {publisher} {Cambridge university Press},\ \bibinfo {year} {2010})\BibitemShut {NoStop}%
\end{thebibliography}%
\end{document}